\RequirePackage{lineno}
 \documentclass[prl,twocolumn,showpacs,superscriptaddress,floatfix]{revtex4-1}
 \def\LineNumbers{false}	
 \def\AuthorList{true} 
 
\usepackage{epsfig}
\usepackage{epstopdf}
\usepackage{graphicx}
\usepackage{float}
\usepackage{amsmath}
\usepackage{bm}
\usepackage{xspace}
\usepackage{lineno}
\usepackage{ifthen}
\usepackage{rotating}
\usepackage{setspace}
\usepackage[abs]{overpic}

\usepackage{ifpdf}
\ifpdf
\else
\fi

\newcommand{\met} {\ensuremath{\mbox{$\protect \raisebox{0.3ex}{$\not$}E_T$}}\xspace}
\newcommand{\metcal} {\ensuremath{\mbox{$\protect \raisebox{0.3ex}{$\not$}E_T$(cal)}}\xspace}

\newcommand{\lv}{\ensuremath{\ell\nu}\xspace}

\newcommand{\bb}{\ensuremath{b\bar{b}}\xspace}
\renewcommand{\tt}{\ensuremath{t\bar{t}}\xspace}
\newcommand{\lvbb}{\ensuremath{\lv\bb}\xspace}
\newcommand{\metbb}{\ensuremath{\met\bb}\xspace}
\newcommand{\gev}{GeV\xspace}

\newcommand{\nnqcd}{\ensuremath{\mathrm{NN}_\mathrm{QCD}}\xspace}

\newcommand{\nntt}{\ensuremath{\mathrm{NN}_{\tt}}\xspace}
\newcommand{\nnnont}{\ensuremath{\mathrm{NN}_{V\mathrm{jets}}}\xspace}
\renewcommand{\nntt}{\ensuremath{\mathrm{NN}_{\tt}}\xspace}
\newcommand{\nnsig}{\ensuremath{\mathrm{NN}_\mathrm{sig}}\xspace}
\newcommand{\tnnsig}{\ensuremath{\nnsig^{t}}\xspace}
\newcommand{\stnnsig}{\ensuremath{\nnsig^{s+t}}\xspace}

\newcommand{\st}{\ensuremath{{s+t}}\xspace}
\newcommand{\pb}{\ensuremath{\mathrm{pb}}\xspace}
\newcommand{\Vtb}{\ensuremath{|V_{tb}|}\xspace}

\begin{document}

\ifthenelse{\equal{\LineNumbers}{true}}{
 \linenumbers }

\title{\boldmath Updated Measurement of the Single Top Quark Production Cross Section and \Vtb in the Missing Transverse Energy Plus Jets Topology in $p\bar{p}$ Collisions at $\sqrt{s} = 1.96$ TeV}

\ifthenelse{\equal{\AuthorList}{true}}{
\affiliation{Institute of Physics, Academia Sinica, Taipei, Taiwan 11529, Republic of China}
\affiliation{Argonne National Laboratory, Argonne, Illinois 60439, USA}
\affiliation{University of Athens, 157 71 Athens, Greece}
\affiliation{Institut de Fisica d'Altes Energies, ICREA, Universitat Autonoma de Barcelona, E-08193, Bellaterra (Barcelona), Spain}
\affiliation{Baylor University, Waco, Texas 76798, USA}
\affiliation{Istituto Nazionale di Fisica Nucleare Bologna, \ensuremath{^{ii}}University of Bologna, I-40127 Bologna, Italy}
\affiliation{University of California, Davis, Davis, California 95616, USA}
\affiliation{University of California, Los Angeles, Los Angeles, California 90024, USA}
\affiliation{Instituto de Fisica de Cantabria, CSIC-University of Cantabria, 39005 Santander, Spain}
\affiliation{Carnegie Mellon University, Pittsburgh, Pennsylvania 15213, USA}
\affiliation{Enrico Fermi Institute, University of Chicago, Chicago, Illinois 60637, USA}
\affiliation{Comenius University, 842 48 Bratislava, Slovakia; Institute of Experimental Physics, 040 01 Kosice, Slovakia}
\affiliation{Joint Institute for Nuclear Research, RU-141980 Dubna, Russia}
\affiliation{Duke University, Durham, North Carolina 27708, USA}
\affiliation{Fermi National Accelerator Laboratory, Batavia, Illinois 60510, USA}
\affiliation{University of Florida, Gainesville, Florida 32611, USA}
\affiliation{Laboratori Nazionali di Frascati, Istituto Nazionale di Fisica Nucleare, I-00044 Frascati, Italy}
\affiliation{University of Geneva, CH-1211 Geneva 4, Switzerland}
\affiliation{Glasgow University, Glasgow G12 8QQ, United Kingdom}
\affiliation{Harvard University, Cambridge, Massachusetts 02138, USA}
\affiliation{Division of High Energy Physics, Department of Physics, University of Helsinki, FIN-00014, Helsinki, Finland; Helsinki Institute of Physics, FIN-00014, Helsinki, Finland}
\affiliation{University of Illinois, Urbana, Illinois 61801, USA}
\affiliation{The Johns Hopkins University, Baltimore, Maryland 21218, USA}
\affiliation{Institut f\"{u}r Experimentelle Kernphysik, Karlsruhe Institute of Technology, D-76131 Karlsruhe, Germany}
\affiliation{Center for High Energy Physics: Kyungpook National University, Daegu 702-701, Korea; Seoul National University, Seoul 151-742, Korea; Sungkyunkwan University, Suwon 440-746, Korea; Korea Institute of Science and Technology Information, Daejeon 305-806, Korea; Chonnam National University, Gwangju 500-757, Korea; Chonbuk National University, Jeonju 561-756, Korea; Ewha Womans University, Seoul, 120-750, Korea}
\affiliation{Ernest Orlando Lawrence Berkeley National Laboratory, Berkeley, California 94720, USA}
\affiliation{University of Liverpool, Liverpool L69 7ZE, United Kingdom}
\affiliation{University College London, London WC1E 6BT, United Kingdom}
\affiliation{Centro de Investigaciones Energeticas Medioambientales y Tecnologicas, E-28040 Madrid, Spain}
\affiliation{Massachusetts Institute of Technology, Cambridge, Massachusetts 02139, USA}
\affiliation{University of Michigan, Ann Arbor, Michigan 48109, USA}
\affiliation{Michigan State University, East Lansing, Michigan 48824, USA}
\affiliation{Institution for Theoretical and Experimental Physics, ITEP, Moscow 117259, Russia}
\affiliation{University of New Mexico, Albuquerque, New Mexico 87131, USA}
\affiliation{The Ohio State University, Columbus, Ohio 43210, USA}
\affiliation{Okayama University, Okayama 700-8530, Japan}
\affiliation{Osaka City University, Osaka 558-8585, Japan}
\affiliation{University of Oxford, Oxford OX1 3RH, United Kingdom}
\affiliation{Istituto Nazionale di Fisica Nucleare, Sezione di Padova, \ensuremath{^{jj}}University of Padova, I-35131 Padova, Italy}
\affiliation{University of Pennsylvania, Philadelphia, Pennsylvania 19104, USA}
\affiliation{Istituto Nazionale di Fisica Nucleare Pisa, \ensuremath{^{kk}}University of Pisa, \ensuremath{^{ll}}University of Siena, \ensuremath{^{mm}}Scuola Normale Superiore, I-56127 Pisa, Italy, \ensuremath{^{nn}}INFN Pavia, I-27100 Pavia, Italy, \ensuremath{^{oo}}University of Pavia, I-27100 Pavia, Italy}
\affiliation{University of Pittsburgh, Pittsburgh, Pennsylvania 15260, USA}
\affiliation{Purdue University, West Lafayette, Indiana 47907, USA}
\affiliation{University of Rochester, Rochester, New York 14627, USA}
\affiliation{The Rockefeller University, New York, New York 10065, USA}
\affiliation{Istituto Nazionale di Fisica Nucleare, Sezione di Roma 1, \ensuremath{^{pp}}Sapienza Universit\`{a} di Roma, I-00185 Roma, Italy}
\affiliation{Mitchell Institute for Fundamental Physics and Astronomy, Texas A\&M University, College Station, Texas 77843, USA}
\affiliation{Istituto Nazionale di Fisica Nucleare Trieste, \ensuremath{^{qq}}Gruppo Collegato di Udine, \ensuremath{^{rr}}University of Udine, I-33100 Udine, Italy, \ensuremath{^{ss}}University of Trieste, I-34127 Trieste, Italy}
\affiliation{University of Tsukuba, Tsukuba, Ibaraki 305, Japan}
\affiliation{Tufts University, Medford, Massachusetts 02155, USA}
\affiliation{University of Virginia, Charlottesville, Virginia 22906, USA}
\affiliation{Waseda University, Tokyo 169, Japan}
\affiliation{Wayne State University, Detroit, Michigan 48201, USA}
\affiliation{University of Wisconsin, Madison, Wisconsin 53706, USA}
\affiliation{Yale University, New Haven, Connecticut 06520, USA}

\author{T.~Aaltonen}
\affiliation{Division of High Energy Physics, Department of Physics, University of Helsinki, FIN-00014, Helsinki, Finland; Helsinki Institute of Physics, FIN-00014, Helsinki, Finland}
\author{S.~Amerio\ensuremath{^{jj}}}
\affiliation{Istituto Nazionale di Fisica Nucleare, Sezione di Padova, \ensuremath{^{jj}}University of Padova, I-35131 Padova, Italy}
\author{D.~Amidei}
\affiliation{University of Michigan, Ann Arbor, Michigan 48109, USA}
\author{A.~Anastassov\ensuremath{^{v}}}
\affiliation{Fermi National Accelerator Laboratory, Batavia, Illinois 60510, USA}
\author{A.~Annovi}
\affiliation{Laboratori Nazionali di Frascati, Istituto Nazionale di Fisica Nucleare, I-00044 Frascati, Italy}
\author{J.~Antos}
\affiliation{Comenius University, 842 48 Bratislava, Slovakia; Institute of Experimental Physics, 040 01 Kosice, Slovakia}
\author{G.~Apollinari}
\affiliation{Fermi National Accelerator Laboratory, Batavia, Illinois 60510, USA}
\author{J.A.~Appel}
\affiliation{Fermi National Accelerator Laboratory, Batavia, Illinois 60510, USA}
\author{T.~Arisawa}
\affiliation{Waseda University, Tokyo 169, Japan}
\author{A.~Artikov}
\affiliation{Joint Institute for Nuclear Research, RU-141980 Dubna, Russia}
% ...

\collaboration{CDF Collaboration}
\altaffiliation[With visitors from]{
\ensuremath{^{a}}University of British Columbia, Vancouver, BC V6T 1Z1, Canada,
\ensuremath{^{b}}Istituto Nazionale di Fisica Nucleare, Sezione di Cagliari, 09042 Monserrato (Cagliari), Italy,
\ensuremath{^{c}}University of California Irvine, Irvine, CA 92697, USA,
\ensuremath{^{d}}Institute of Physics, Academy of Sciences of the Czech Republic, 182~21, Czech Republic,
\ensuremath{^{e}}CERN, CH-1211 Geneva, Switzerland,
\ensuremath{^{f}}Cornell University, Ithaca, NY 14853, USA,
\ensuremath{^{g}}University of Cyprus, Nicosia CY-1678, Cyprus,
\ensuremath{^{h}}Office of Science, U.S. Department of Energy, Washington, DC 20585, USA,
\ensuremath{^{i}}University College Dublin, Dublin 4, Ireland,
\ensuremath{^{j}}ETH, 8092 Z\"{u}rich, Switzerland,
\ensuremath{^{k}}University of Fukui, Fukui City, Fukui Prefecture, Japan 910-0017,
\ensuremath{^{l}}Universidad Iberoamericana, Lomas de Santa Fe, M\'{e}xico, C.P. 01219, Distrito Federal,
\ensuremath{^{m}}University of Iowa, Iowa City, IA 52242, USA,
\ensuremath{^{n}}Kinki University, Higashi-Osaka City, Japan 577-8502,
\ensuremath{^{o}}Kansas State University, Manhattan, KS 66506, USA,
\ensuremath{^{p}}Brookhaven National Laboratory, Upton, NY 11973, USA,
\ensuremath{^{q}}Queen Mary, University of London, London, E1 4NS, United Kingdom,
\ensuremath{^{r}}University of Melbourne, Victoria 3010, Australia,
\ensuremath{^{s}}Muons, Inc., Batavia, IL 60510, USA,
\ensuremath{^{t}}Nagasaki Institute of Applied Science, Nagasaki 851-0193, Japan,
\ensuremath{^{u}}National Research Nuclear University, Moscow 115409, Russia,
\ensuremath{^{v}}Northwestern University, Evanston, IL 60208, USA,
\ensuremath{^{w}}University of Notre Dame, Notre Dame, IN 46556, USA,
\ensuremath{^{x}}Universidad de Oviedo, E-33007 Oviedo, Spain,
\ensuremath{^{y}}CNRS-IN2P3, Paris, F-75205 France,
\ensuremath{^{z}}Universidad Tecnica Federico Santa Maria, 110v Valparaiso, Chile,
\ensuremath{^{aa}}The University of Jordan, Amman 11942, Jordan,
\ensuremath{^{bb}}Universite catholique de Louvain, 1348 Louvain-La-Neuve, Belgium,
\ensuremath{^{cc}}University of Z\"{u}rich, 8006 Z\"{u}rich, Switzerland,
\ensuremath{^{dd}}Massachusetts General Hospital, Boston, MA 02114 USA,
\ensuremath{^{ee}}Harvard Medical School, Boston, MA 02114 USA,
\ensuremath{^{ff}}Hampton University, Hampton, VA 23668, USA,
\ensuremath{^{gg}}Los Alamos National Laboratory, Los Alamos, NM 87544, USA,
\ensuremath{^{hh}}Universit\`{a} degli Studi di Napoli Federico I, I-80138 Napoli, Italy
}
\noaffiliation
 }

\ifthenelse{\equal{\AuthorList}{false}}{
  \begin{figure}  
    \leftline{\includegraphics[scale=0.5]{images/cdfii_thumb_logo}\hfill
      CDF/PUB/TOP/CDFR/11033}
  \end{figure}
  
  \vspace*{0.5cm} 
  
  \author{The CDF Collaboration}
  \affiliation{URL http://www-cdf.fnal.gov}
}

\date{\today}

\begin{abstract}
An updated measurement of the single top quark production cross
section is presented using the full data set collected by the Collider
Detector at Fermilab (CDF) and corresponding to 9.5 fb${}^{-1}$ of
integrated luminosity from proton-antiproton collisions at 1.96 TeV
center-of-mass energy.  The events selected contain an imbalance in
the total transverse energy, jets identified as originating from $b$
quarks, and no identified leptons.  The sum of the $s$- and
$t$-channel single top quark cross sections is measured to be
$3.53_{-1.16}^{+1.25}$ pb and a lower limit on \Vtb of 0.63 is
obtained at the 95\% credibility level.  These measurements are
combined with previously reported CDF results obtained from events
with an imbalance in total transverse energy, jets identified as
originating from $b$ quarks, and exactly one identified lepton.  The
combined cross section is measured to be $3.02_{-0.48}^{+0.49}$ pb and
a lower limit on \Vtb of 0.84 is obtained at the 95\% credibility
level.
\end{abstract}


\maketitle


The observation of single top quark production at the Tevatron was a
significant achievement, allowing measurements of the cross section at
a hadron collider~\cite{stopDisc} and improved bounds on the
Cabibbo-Kobayashi-Maskawa (CKM)~\cite{ckm} matrix element magnitude
\Vtb due to the direct coupling of the $b$ quark with the singly
produced top quark.

For single top quark production, a $t$-$\bar{b}$ pair is produced
through a virtual $W^+$ boson~\cite{charge} in either the $s$ or $t$
channel. The top quark subsequently decays to a $W^+$ boson and a
bottom quark, and the fragmentations of the $b$ and $\bar{b}$ quarks
result in two jets that can be reconstructed in the detector.  For the
$t$-channel process, jets tend to be more boosted along the
proton-antiproton beam axis than those originating from the
$s$-channel process; some of these jets are thus emitted in regions
that are not instrumented and therefore escape the detector
acceptance.

Excluding the contribution from the $tW$ production mode, which is
expected to be negligible in the final state considered in this
Letter~\cite{tw}, the standard model (SM) prediction for the combined
$s$- and $t$-channel single top quark production cross section
$\sigma^\st_\mathrm{SM}$ is $3.15 \pm 0.36~\pb$, which has been
calculated including next-to-next-to-leading order
corrections~\cite{kidonakis,ststbar}.  The primary sensitivity to
measuring this quantity is usually obtained from events where the $W$
boson from the $t \to W b$ process~\cite{twb} decays leptonically to a
charged lepton $\ell$ (where $\ell$ represents either an electron $e$ or muon
$\mu$) and an antineutrino, with a pair of jets, one of which is
``$b$-tagged'' or identified as likely having originated from a bottom
quark.  This sample of events (hereafter the ``\lvbb'' sample)
provides a distinctive signature against backgrounds produced by the
strong interaction (QCD multijet or ``MJ'' background), which contain
no leptons and multiple jets.

A complementary approach consists in using final states that contain
two or three jets and significant imbalance in the total transverse
energy \met~\cite{MET}, which results from the leptonic decay of the
$W$ boson, where the lepton is not identified due to reconstruction or
acceptance effects and the neutrino carries significant unmeasured
momentum.  Although MJ events comprise the dominant background in this
final state (hereafter the ``\metbb'' analysis or sample), the
requirement of significant $\met$ greatly suppresses such background.
In addition, this search has sensitivity to events where the $W$ boson
decays via $W^-\to\tau^-\bar{\nu}_\tau$, and the $\tau^-$ decays
hadronically, resulting in a reconstructed jet signature.

The first CDF measurement of single top quark production in the $\met
b\bar{b}$ final state was performed with a data set corresponding to
an integrated luminosity of 2.1~$\mathrm{fb}^{-1}$~\cite{metJetsPRD}.
This article presents a new measurement using the full CDF data set
(9.5~$\mathrm{fb}^{-1}$). All the techniques developed in the search
for $s$-channel single top quark production in the \metbb
sample~\cite{metJetsSch} are exploited in this update. Important
aspects of the analysis methodology are restated here for
completeness.  The results of this analysis and those of the most
recent $\ell \nu b\bar{b}$ analysis~\cite{lvbbSch} are then combined
to obtain a more precise measurement of the single top quark cross
section and to place a lower limit on the CKM matrix element magnitude
\Vtb.


The CDF II detector is a multipurpose particle detector described in
detail elsewhere~\cite{cdf}.  It is comprised of an inner silicon
vertex detector, a 96-layer drift chamber spectrometer used for
reconstructing charged-particle trajectories (tracks), and a
calorimeter that is divided radially into electromagnetic and hadronic
compartments, which are constructed of projective towers that cover
pseudorapidities of up to $|\eta| < 3.6$~\cite{geometry}.  A system of
drift chambers located outside the hadronic calorimeter is used for
muon identification.

Jets are formed by clustering calorimeter energy deposits with an
opening angle of $\Delta R \equiv \sqrt{(\Delta
\eta)^2+(\Delta\phi)^2} = 0.4$.  Lepton candidates with large transverse 
momentum are identified by associating tracks with signatures in the
appropriate detectors: energy deposits in the electromagnetic
calorimeters for electrons, and muon-detector track segments for
muons.

Events are selected in which the calorimeter missing transverse energy
\metcal satisfies a minimum online selection (trigger) threshold of at least
45~\gev, or 35~\gev if at least two jets are present.  In the offline
analysis, events are accepted if the reconstructed missing transverse
energy \met is at least 35~\gev.  A multivariate algorithm is used to
parameterize the trigger efficiency as a function of several kinematic
and angular variables of the event~\cite{karolos}.  Measured jet
energies are corrected to account for irregularities in calorimeter
response, energy lost outside the jet cone, and underlying event
dynamics~\cite{bhatti06}.  The jet energy scale and resolution, as
well as the \met resolution, are further improved by incorporating
corrections based on charged-particle momentum
measurements~\cite{adloff97}.

Each event is required to have at least two leading-$E_T$ jets with
transverse energies, $E_T^{j_1}$ and $E_T^{j_2}$, that satisfy $25\ <\
E_T^{j_1}\ <\ 200$ GeV and $20\ <\ E_T^{j_2}\ <\ 120$ GeV,
respectively.  Additionally, both leading-$E_T$ jets are required to
be reconstructed within the silicon detector acceptance, corresponding
to pseudorapidity requirements of $|\eta|<2$ for both jets, with one
of them satisfying $|\eta|<0.9$. Events with three jets are considered
if the third-most energetic jet in $E_T$ satisfies $15\ <\ E_T^{j_3}\
<\ 100$ GeV and $|\eta| <2.4$. Events with four or more jets are
rejected if each jet satisfies the criteria $E_T > 15$ GeV and $|\eta|
< 2.4$.  To discriminate against MJ background, the angular separation
between the two highest-$E_T$ jets must satisfy $\Delta R > 0.8$.
Events that satisfy these requirements are labeled ``pre-tagged''
events.

To suppress light-flavor MJ background, at least one of the
leading-$E_T$ jets is required to be $b$-tagged by the \textsc{hobit}
algorithm~\cite{hobit}, which assigns to each jet a value between 0
and 1.  Jets with a \textsc{hobit} value between 0.72 (0.95) and 0.95
(1) are considered to be loosely (tightly) tagged.  As two $b$ quarks
are present in the signal final state, events are separated into three
categories based on the multiplicity and quality of the $b$-tagged
jets: events with only one tightly tagged jet and no other tag (1T),
events with two tightly tagged jets (TT), and events with one tightly
tagged jet and one loosely tagged jet (TL). Events are further
classified according to the total number of jets, leading to six event
subsamples.  Each subsample is analyzed separately to improve the
sensitivity and to help separate the $s$- and $t$-channel produced
events, which are enhanced in the double- and single-tagging
categories, respectively.

All events that satisfy the above kinematic and $b$-tagging criteria
are separated into two samples.  Events that contain no identified
leptons comprise the preselection sample, which includes events in the
signal region, defined below.  Events that contain at least one
identified electron or muon comprise the electroweak sample, which is
used to validate the background modeling derived for this analysis.

Most physics processes are modeled using Monte Carlo simulation
programs.  The single top quark samples are modeled using the
\textsc{powheg} generator~\cite{powheg}.  Backgrounds from $V$+jets
(where $V$ represents a $W$ or $Z$ boson) and $W+c$ processes are
modeled using \textsc{alpgen}~\cite{alpgen}, with showering simulated
by \textsc{pythia}~\cite{pythia}.  Events from diboson ($VV$),
$t\bar{t}$ (assuming a top-quark mass of 172.5~\gev/$c^2$), and Higgs
bosons produced in association with a $W$ or $Z$ boson ($V\!H$) are
simulated using \textsc{pythia}.  Two remaining backgrounds include
contributions from events with falsely tagged jets (``electroweak
mistags'') and MJ events.

The electroweak mistag samples are modeled by weighting $V$+jets and
diboson-simulated events with mistag probabilities derived from
dedicated data samples~\cite{secvtx,mistag}.  To model the MJ
background, the same data-driven method described in
Ref.~\cite{higgsmetbb} is used: each pretagged data event is weighted
by a tag-rate probability derived from a MJ-dominated data sample.

\begin{figure*}[htbp]
    \includegraphics[width=0.3\linewidth]{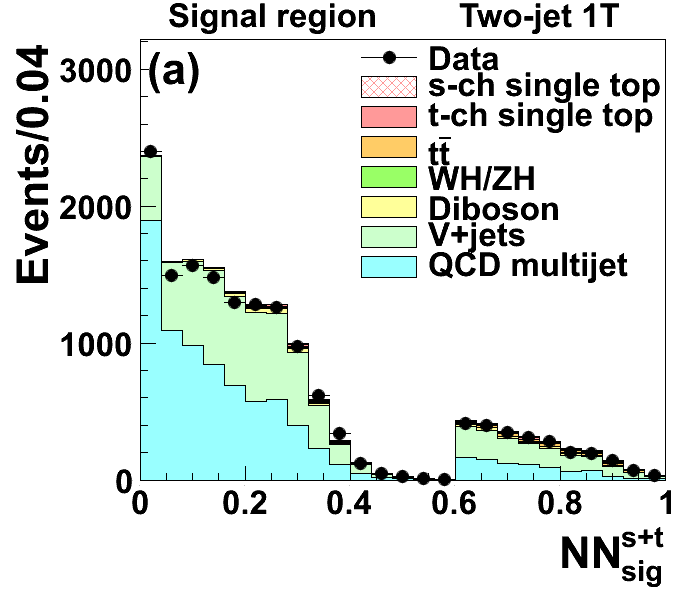} \hfill
    \includegraphics[width=0.3\linewidth]{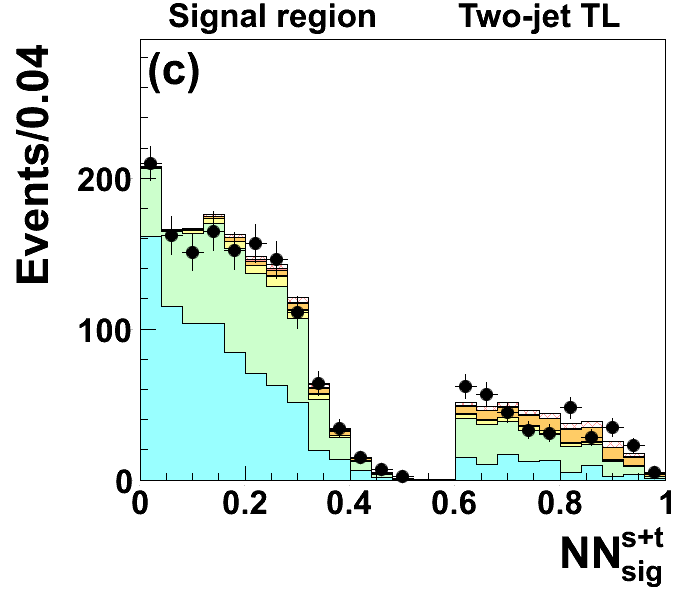} \hfill
    \includegraphics[width=0.3\linewidth]{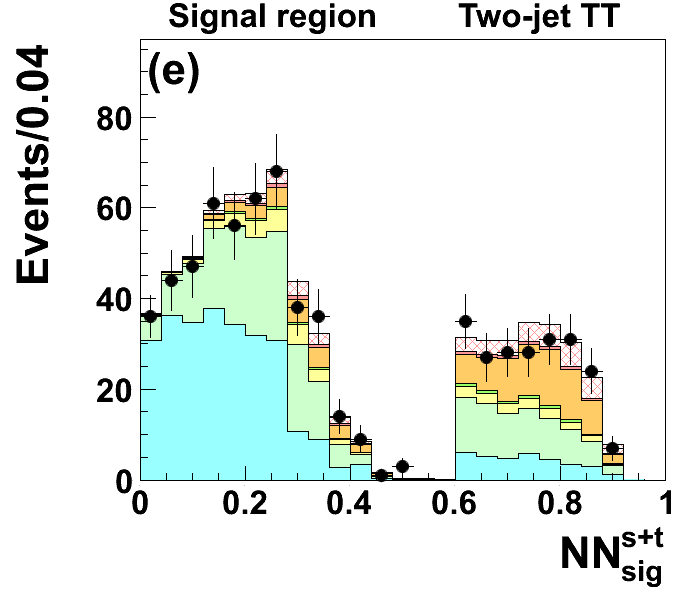} \\
    \includegraphics[width=0.3\linewidth]{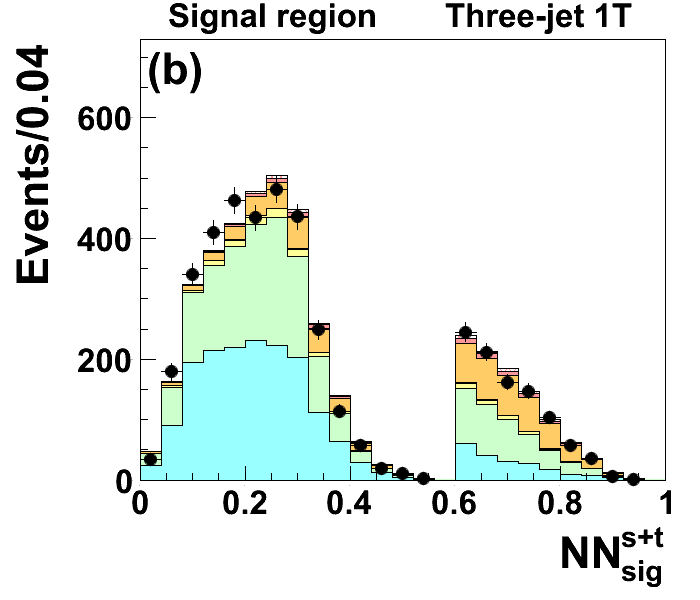} \hfill
    \includegraphics[width=0.3\linewidth]{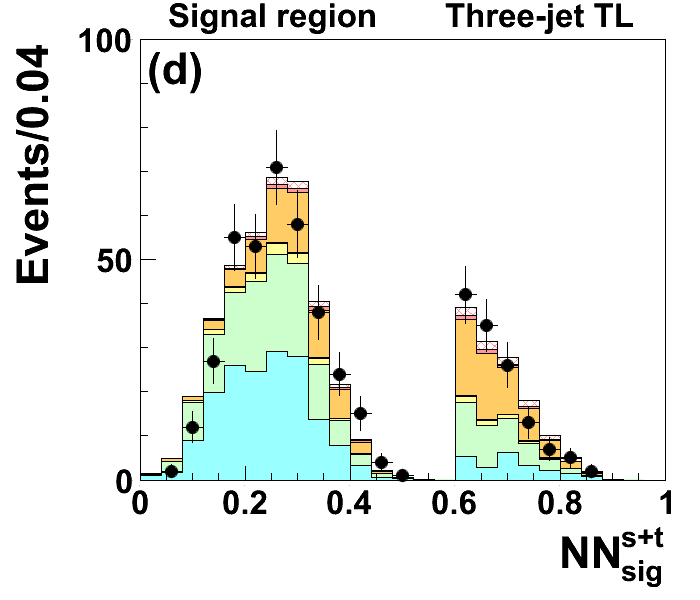} \hfill
    \includegraphics[width=0.3\linewidth]{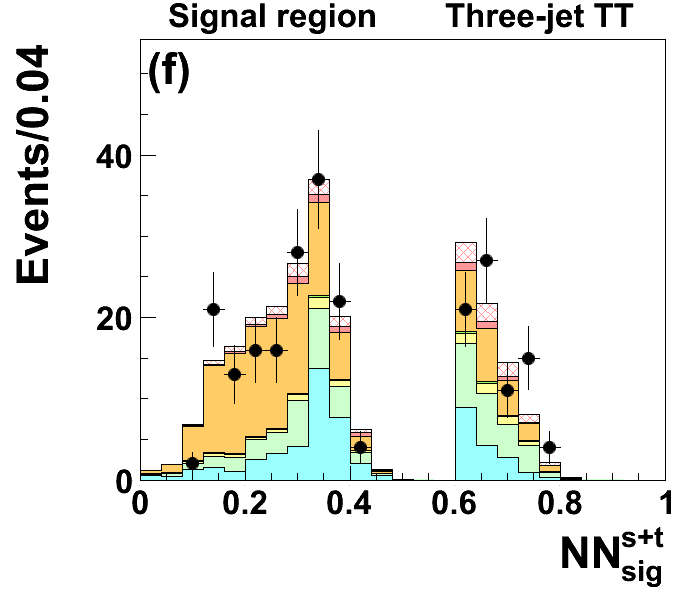}
    \caption{ Predicted and observed \stnnsig\ distributions in the signal region,
for the (a) 1T two-jet, (b) 1T three-jet, (c) TL two-jet, (d) TL
three-jet, (e) TT two-jet, (f) and TT three-jet
subsamples.}
\label{fig:stnnsig}
\end{figure*}

At this stage of the analysis, simple requirements on kinematic
properties of the event are not sufficient to separate the single top
quark signal from the background.  A series of multivariate
discriminants that take advantage of nontrivial variable correlations
are therefore employed to optimize the suppression of MJ background
and to separate the signal from the remaining backgrounds.  For each
of the multivariate algorithms described below, a combination of
inputs is used corresponding to kinematic, angular, and event-topology
quantities whose distributions are different between the background
under consideration and the single top quark signal.

The dominant background in the preselection sample is MJ events.  To
discriminate against this background, the same \nnqcd multivariate
discriminant that was developed in the $\met b\bar{b}$ $s$-channel
single top quark search~\cite{metJetsSch} is used.  All events that
satisfy a minimum \nnqcd threshold requirement populate the signal
region, in which the dominant backgrounds are from MJ production,
$V$+heavy-flavor-jets events, and $t\bar{t}$ events.  Events that do
not meet the minimal \nnqcd threshold are used to validate the
background prediction with the data. From this validation,
multiplicative correction factors ranging from 0.7 to 0.9 are derived
for each of the 1T, TL and TT MJ predictions so that the total
predicted background normalizations are in agreement with data.  These
corrections are applied to the MJ predictions in the signal region.

For all events in the signal region, two additional discriminants are
developed that further exploit the differences in kinematic properties
between the signal and the $V$+jets background, and the signal and
$t\bar{t}$ background processes.  The first discriminant \nnnont is
trained using simulated $t$-channel single top quark events for the
signal sample and MJ-modeled events that satisfy the requirement on
\nnqcd, for the background sample. The second discriminant \nntt is
trained to separate $t$-channel single top quark from $t\bar{t}$
production, again using simulated $t$-channel single top quark events
for the signal but simulated $t\bar{t}$ for the background.  The
values of these two discriminants (both supporting values between 0
and 1) are then combined in quadrature for an overall discriminant
called \tnnsig; this is analogous to the strategy adopted to derive
the final discriminant in the \metbb $s$-channel single top quark
search~\cite{metJetsSch}.

The $s$-channel optimized \nnsig discriminant as used in the \metbb
$s$-channel single top quark search~\cite{metJetsSch} and the
\tnnsig discriminant of this analysis are combined to obtain an 
\stnnsig final discriminant, used to simultaneously separate both $s$- and $t$-channel signal
processes from the remaining background. For events with \nnsig output
values larger than 0.6, \stnnsig is assigned to the \nnsig output. For
the remaining events, \stnnsig is defined as the \tnnsig output
multiplied by 0.6. Figure~\ref{fig:stnnsig} shows the predicted and
observed distributions of the \stnnsig\ output variable for each of
the six event subsamples used in this analysis.

Several sources of systematic uncertainty are taken into
account. Uncertainties on assumed cross section values are included
for $t\bar{t}$ (5\%-6\%), $VV$ (6\%), $V\!H$ (5\%), and $W+c$ (23\%)
processes~\cite{top,diboson,hbb_prod,wc}.  Other systematic
uncertainties arise from the normalization of the
$V$-plus-heavy-flavor (30\%) and of the MJ (3\%-7\%) background
contributions.  All samples whose normalizations are not constrained
according to the data are subject to a luminosity uncertainty of
6\%~\cite{lumi_cdf}. Furthermore, uncertainties are assigned due to
the efficiencies of the lepton antiselection criteria (2\%).  We also
assign a normalization uncertainty of 2\% due to variations in the
assumed parton distribution functions.  To account for differences in
the trigger efficiency in data and simulation, a 2\% rate uncertainty
is assigned.

Possible mismodeling in the $b$-tagging efficiency is taken into
account by applying scale factors to the simulation to correct its
$b$-tagging efficiency to that of data.  All scale factors are
determined from data control samples and are on the order of unity,
with small variations depending on the tag category, and uncertainties
ranging from 8\% to 16\%~\cite{hobit}. Mistag rate uncertainties
(20\%-30\%) are also derived from data and included~\cite{hobit}.

Uncertainties in the jet energy scale~\cite{bhatti06} are included by
correlating the uncertainties in the predicted yields of signals and
backgrounds (of the order of 1\%-6\%) with the corresponding
distortions in the predicted kinematic distributions arising from jet
energy scale shifts in all samples except the MJ background, which is
determined entirely from data.  An additional systematic uncertainty
is incorporated for the MJ model, accounting for shape variations in
the MJ prediction.
                                                                       
To measure the signal cross section, a combined likelihood is formed,
which is the product of Poisson probabilities for each bin of the six
\stnnsig discriminants shown in Fig.~\ref{fig:stnnsig}.  To account
for systematic uncertainties, a Bayesian technique is used, in which
each independent source of systematic uncertainty is assigned a
nuisance parameter with a Gaussian prior probability density,
truncated when necessary to ensure non-negative event yields.  The
impact of each nuisance parameter is propagated to the predictions of
the signal and background yields in each bin of each histogram in the
analysis.  A non-negative uniform prior probability distribution is
assumed for the single top quark cross section, which is extracted
from its posterior probability density after integrating over all
nuisance parameters.

\begin{figure*}[htbp]
    \begin{overpic}[width=0.48\textwidth]{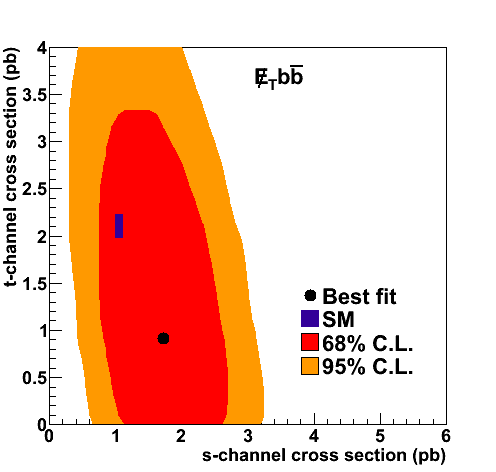}
    \ifthenelse{\equal{\LineNumbers}{true}}{ \put(150,176){\bf(a)} }{ \put(200,191){\bf(a)} }
    \end{overpic} \hfill
    \begin{overpic}[width=0.48\textwidth]{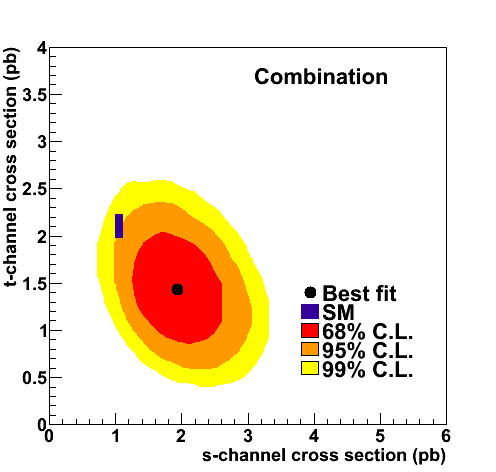}
    \ifthenelse{\equal{\LineNumbers}{true}}{ \put(150,177){\bf(b)} }{ \put(200,192){\bf(b)} }
    \end{overpic}
    \caption{Two-dimensional posterior probability densities of the
    $s$- and $t$-channel cross sections for the (a) \metbb analysis
    presented here, and (b) the CDF combination of the \metbb and \lvbb
    analysis results.  Theory uncertainties are not shown.}  
    \label{fig:twod}
\end{figure*}

Tables~\ref{sig2} and~\ref{sig3} show the event yields in the two- and
three-jet subsamples, respectively, as determined from applying the
measurement procedure to the six discriminants shown in
Fig.~\ref{fig:stnnsig}. The observed single top quark production cross
section ($\sigma_\mathrm{obs}^{\st}$) is $3.53_{-1.16}^{+1.25}$ pb,
consistent with the SM prediction of $3.15 \pm 0.36$
pb~\cite{kidonakis}.  The magnitude of $V_{tb}$ is extracted from the
single top quark cross section posterior probability density by the
relation
$\Vtb^{2}_\mathrm{obs}=\Vtb^{2}_\mathrm{SM}\sigma^{\st}_\mathrm{obs}/\sigma^{\st}_\mathrm{SM}$,
where variables with the subscript ``SM'' (``obs'') correspond to the
theoretical predictions (observed values)~\cite{vtb}.  We assume
$\Vtb^{2}_\mathrm{SM}$ is unity and fix the $s$- and $t$-channel
relative contributions to their SM prediction. Including the
theoretical uncertainty of the signal cross section (5.8\% for
$s$-channel, 6.2\% for $t$-channel)~\cite{kidonakis} and assuming a
uniform prior in the interval $0\ <\ \Vtb^2\ <\ 1$, a lower bound on
\Vtb of 0.63 is obtained at the 95\% credibility level
(C.L.).  We measure the $t$-channel cross section by itself to be
$1.19^{+0.93}_{-0.97}$ pb, where the $s$-channel cross section is
constrained to its SM prediction~\cite{schannelnote}; this result is
consistent with the SM prediction of $2.10 \pm 0.12$
pb~\cite{kidonakis}.  We also compute the two-dimensional posterior
for the $s$- and $t$-channel cross sections, where the relative
contributions of both channels are allowed to vary independently; the
result is shown in Fig.~\ref{fig:twod}a.

\begin{table}[tb]
  \begin{center}

    \caption{\label{sig2}Numbers of predicted and observed events in
    the two-jet signal region in the subsamples with exactly one
    tightly tagged jet (1T), one tightly and one loosely tagged jet
    (TL), and two tightly tagged jets (TT). The uncertainties in the
    predicted numbers of events are due to the theoretical cross
    section uncertainties and to the uncertainties on signal and
    background modeling. Both the uncertainties and the central values
    are those determined by the fit to the data with theory constraints.}

\scriptsize
  \begin{tabular}{c*{3}{c}}
     \hline
     \hline
     Category                 & 1T & TL & TT \\
     \hline
     $t\bar{t}$ &    242.9 $\pm$   24.3 &     84.8 $\pm$    9.3 &     92.4 $\pm$    8.4\\ 
           $VH$ &     12.6 $\pm$    1.4 &      6.6 $\pm$    0.8 &      7.6 $\pm$    0.8\\ 
        Diboson &    284.9 $\pm$   25.6 &     51.3 $\pm$    4.6 &     37.2 $\pm$    3.4\\ 
       $V$+jets &   6527.7 $\pm$ 2048.1 &    694.2 $\pm$  216.3 &    220.2 $\pm$   68.7\\ 
             MJ &   8328.5 $\pm$  180.6 &    885.2 $\pm$   56.7 &    296.8 $\pm$   31.8\\ 
     \hline	        
$s$-ch single top &     86.2 $\pm$   47.7 &     41.8 $\pm$   23.2 &     45.9 $\pm$   25.3\\ 
$t$-ch single top &    160.5 $\pm$   30.8 &     10.8 $\pm$    2.1 &      9.2 $\pm$    1.7\\ 

     \hline
     Total prediction   &  15643.4 $\pm$ 2057.2 &   1774.8 $\pm$  225.0 &    709.3 $\pm$   80.3\\
     \hline
     Observed    &                   15312  &             1743  &               686 \\
     \hline
     \hline
    \end{tabular}        
  \end{center}
\end{table}

\begin{table}[tb]
  \begin{center}

   \caption{\label{sig3}Numbers of predicted and observed three-jet events in the 1T, TL, and TT subsamples.}

\scriptsize
  \begin{tabular}{c*{3}{c}}
    	\hline
      \hline
     Category                 & 1T & TL & TT \\
     \hline
     $t\bar{t}$ &    596.5 $\pm$   59.6 &    117.5 $\pm$   12.8 &    109.5 $\pm$    9.9\\ 
           $VH$ &      6.0 $\pm$    0.7 &      1.9 $\pm$    0.2 &      2.2 $\pm$    0.2\\ 
        Diboson &    107.7 $\pm$    9.7 &     15.7 $\pm$    1.5 &      8.8 $\pm$    0.8\\ 
       $V$+jets &   1609.5 $\pm$  505.1 &    164.5 $\pm$   51.3 &     50.4 $\pm$   15.8\\ 
             MJ &   1818.2 $\pm$   48.7 &    187.5 $\pm$   14.7 &     55.9 $\pm$    7.7\\ 
     \hline
$s$-ch single top &     45.7 $\pm$   25.3 &     15.4 $\pm$    8.5 &     16.2 $\pm$    8.9\\ 
$t$-ch single top &     82.2 $\pm$   15.8 &      7.5 $\pm$    1.5 &      6.8 $\pm$    1.3\\ 
     \hline
     Total prediction    &   4265.7 $\pm$  511.9 &    510.0 $\pm$   55.6 &    249.8 $\pm$   22.1\\
     \hline
     Observed    &              4198  &                  490  &                  237 \\
     \hline
     \hline
    \end{tabular}        
  \end{center}
\end{table}

These results are combined with those of the most recent CDF
measurement of single top quark production in the \lvbb
sample~\cite{lvbbSch}, which measured a cross section of
$3.04^{+0.57}_{-0.53}$ pb assuming a top quark mass of
172.5~\gev/$c^2$. The combination is achieved by taking the product of
the likelihoods of both analyses and simultaneously varying the
correlated uncertainties, following the procedure explained above. In
the \lvbb analysis, candidate events were selected by requiring
exactly one reconstructed charged lepton ($e$ or $\mu$) in the final
state. Hence, no such events are included in the \metbb analysis
described above. The uncertainties associated with the theoretical
cross sections of the $t\bar{t}$, $VV$, and $V\!H$ production
processes, and those associated with the luminosity are taken as fully
correlated between the two analyses.

The combined measurement results in an electroweak single top quark
production cross section of $3.02_{-0.48}^{+0.49}$ pb, consistent with
the SM prediction.  From the posterior probability density on
$\Vtb^2$, a 95\% C.L. lower limit of $\Vtb > 0.84$ is obtained. The
$t$-channel cross section, measured in the same way as for the \metbb
analysis, is $1.65^{+0.38}_{-0.36}$ pb, in agreement with the SM
prediction given above.  The two-dimensional posterior probability is
shown in Fig.~\ref{fig:twod}b, where the relative $s$- and $t$-channel
contributions are allowed to vary freely.

\begin{figure*}[htbp]

\end{figure*}

In summary, an updated measurement of the single top quark production
cross section of $3.53_{-1.16}^{+1.25}$ pb is obtained in events with
missing transverse energy and jets using the full CDF data set.  This
represents a relative improvement of 40\% in overall precision with
respect to the previous CDF analysis~\cite{metJetsPRD}.  In addition,
a combination with the \lvbb CDF result~\cite{lvbbSch} is performed to
obtain a more precise cross section measurement of
$3.04^{+0.57}_{-0.53}$ pb.  Cross sections for the $t$-channel-only
single top quark production process as well as 95\% C.L. lower limits
on \Vtb are also obtained for the \metbb analysis and the CDF
combination.  All results are consistent with the corresponding SM
predictions.


We thank the Fermilab staff and the technical staffs of the
participating institutions for their vital contributions. This work
was supported by the U.S. Department of Energy and National Science
Foundation; the Italian Istituto Nazionale di Fisica Nucleare; the
Ministry of Education, Culture, Sports, Science and Technology of
Japan; the Natural Sciences and Engineering Research Council of
Canada; the National Science Council of the Republic of China; the
Swiss National Science Foundation; the A.P. Sloan Foundation; the
Bundesministerium f\"ur Bildung und Forschung, Germany; the Korean
World Class University Program, the National Research Foundation of
Korea; the Science and Technology Facilities Council and the Royal
Society, United Kingdom; the Russian Foundation for Basic Research;
the Ministerio de Ciencia e Innovaci\'{o}n, and Programa
Consolider-Ingenio 2010, Spain; the Slovak R\&D Agency; the Academy
of Finland; the Australian Research Council (ARC); and the EU community
Marie Curie Fellowship Contract No. 302103.


\bibliography{singleTop_metJets}

\begin{thebibliography}{30}
\bibitem[{cha()}]{charge}
\bibinfo{note}{Charge-conjugate processes are assumed throughout unless
  explicitly stated otherwise.}
\bibitem[{twb()}]{twb}
\bibinfo{note}{We assume a branching ratio of 100\% for the standard-model $t
  \to W^+ b$ process.}
\bibitem[{MET()}]{MET}
\bibinfo{note}{The calorimeter missing transverse energy $\vmet(\mbox{cal})$ is
  defined by the sum over calorimeter towers, $\vmet(\mbox{cal}) = - \sum_{i}
  E_T^i \hat{n}_i$, where $i$ is a calorimeter tower number with $|\eta| <
  3.6$, and $\hat{n}_i$ is a unit vector perpendicular to the beam axis and
  pointing at the $i$th calorimeter tower. The reconstructed missing transverse
  energy, $\vmet$, is derived by subtracting from $\vmet(\mbox{cal})$
  components of the event not registered by the calorimeter, such as jet energy
  adjustments. $\met$ [$\met(\mbox{cal})$] is the scalar magnitude of $\vmet$
  [$\vmet(\mbox{cal})$].}
\bibitem[{geo()}]{geometry}
\bibinfo{note}{We use a cylindrical coordinate system where $\theta$ is the
  polar angle relative to the proton beam direction at the event vertex, $\phi$
  is the azimuthal angle about the beam axis, and pseudorapidity is defined
  $\eta = -\ln\tan(\theta/2)$. We define transverse energy as $E_T =
  E\sin\theta$ and transverse momentum as $p_T = p\sin\theta$ where $E$ is the
  energy measured in the calorimeter and $p$ is the magnitude of the momentum
  measured by the spectrometer.}
\bibitem[{vtb()}]{vtb}
\bibinfo{note}{This relation is robust assuming that $\Vtb^2 \gg |V_{ts}|^2 +
  |V_{td}|^2$.}
\bibitem[{sch()}]{schannelnote}
\bibinfo{note}{This Letter does not report a separate $s$-channel cross section
  result, which is presented in Ref.~\cite{metJetsSch}.}
\end{thebibliography}

\end{document}